\documentclass[10pt]{iopart}

\usepackage[backend=biber, style=phys]{biblatex}  %
\expandafter\let\csname equation*\endcsname\relax
\expandafter\let\csname endequation*\endcsname\relax
\usepackage{amsmath, amsbsy}
\usepackage{graphicx}
\usepackage[unicode=true, bookmarks=true, bookmarksnumbered=true, bookmarksopen=true, bookmarksopenlevel=1, breaklinks=false,pdfborder={0 0 0}, backref=false, colorlinks=true]{hyperref}
\hypersetup{citecolor=blue,linkcolor=blue}
\usepackage{breakurl}

\usepackage[dvipsnames]{xcolor} %
\usepackage{lipsum,multicol} %

\usepackage [english]{babel}
\usepackage [autostyle, english = american]{csquotes}
\MakeOuterQuote{"}

\usepackage{subcaption}

\usepackage{graphicx}

\usepackage[final]{changes}

\begin{document}

\title{Error Field Predictability and Consequences for ITER}

\author{M. Pharr$^1$
N.C. Logan$^1$,
C. Paz-Soldan$^1$,
J.K. Park$^{2,3}$,
C. Hansen$^1$
}
\address{
    $^1$ Columbia University, New York, New York 10027, USA
}
\address{
    $^2$ Seoul National University, Seoul 08826, Republic of Korea
}
\address{
    $^3$ Princeton Plasma Physics Laboratory, Princeton, New Jersey 08540, USA
}
 \ead{matthew.pharr@columbia.edu}
 
\vspace{10pt}
\begin{indented}
\item[]\today
\end{indented}

\begin{abstract}

ITER coil tolerances are re-evaluated using the modern understanding of coupling to least-stable plasma modes and an updated center-line-traced model of ITER's coil windings. This reassessment finds the tolerances to be conservative through a statistical, linear study of $n=1$ error fields (EFs) due to tilted, shifted misplacements and nominal windings of central solenoid and poloidal field coils within tolerance. We also show that a model-based correction scheme remains effective even when metrology quality is sub-optimal, and compare this to projected empirical correction schemes. We begin with an analysis of the necessity of error field correction (EFC) for daily operation in ITER using scaling laws for the EF penetration threshold. We then consider the predictability of EF dominant mode overlap across early planned ITER scenarios and, as measuring EFs in high power scenarios can pose risks to the device, the potential for extrapolation to the ITER Baseline Scenario (IBS). We find that carefully designing a scenario matching currents proportionally to those of the IBS is far more important than plasma shape or profiles in accurately measuring an optimal correction current set.

\end{abstract}

\maketitle
\ioptwocol

\section{\label{sec:Introduction} Introduction}

Non-axisymmetric magnetic fields are known to degrade tokamak plasma performance and destabilize magnetohydrodynamic (MHD) modes. These fields have been shown to degrade pedestal confinement \cite{paz-soldan_decoupled_2015, park_mdc-19_2017}, to damp plasma rotation through Neoclassical Toroidal Viscosity (NTV) \cite{shaing_symmetry-breaking_2001,shaing_neoclassical_2015}, and to resonate at rational surfaces with MHD modes \cite{fitzpatrick_interaction_1993, buttery_error_1999, buttery_error_2000, buttery_impact_2011, hender_effect_1992,karger_influence_1974}. Error fields due to a device's inherent non-axisymmetric properties such as coil windings, inexact coil placement, coil deformation, ferritic structural elements, coil leads, and other sources, also known as intrinsic error fields, can cause mode locking and subsequent disruption at amplitudes as small as $10^{-4} B_T$ \cite{buttery_error_2000}. 

The International Tokamak Physics Activity (ITPA) advising the planning of ITER has used several metrics to quantify error fields. The previous metric was known as the 3-mode representation \cite{scoville_multi-mode_2003, hender_chapter_2007, amoskov_fourier_2004, amoskov_statistical_2005}; this was a weighted geometric average of the first three ($m=1,2,3$) Fourier harmonics of the radial magnetic \emph{field} calculated in vacuum. References \cite{schaffer_study_2008, park_spectral_2008} showed that these harmonics are coordinate-system dependent. Alongside experimental evidence \cite{paz-soldan_spectral_2014, paz-soldan_importance_2014} for the trouble with the 3-mode metric, this was sufficient to motivate the transition to a metric known as the resonant mode overlap. This metric employs both Fourier harmonics of the magnetic flux, which are half-area weighted to be coordinate system independent, and a singular value decomposition (SVD) technique to choose the weightings of contributions from different harmonics. The latter is an advantage over using the 2/1 flux, as the overlap can capture the effects of locking risk at 3/1 and higher surfaces. This metric is now favored by the ITPA and is now the predominant metric for identifying error fields in tokamaks; it has been the subject of numerous studies on a range of devices, including DIII-D \cite{paz-soldan_non-disruptive_2022,paz-soldan_importance_2014,paz-soldan_spectral_2014,park_error_2011}, COMPASS/COMPASS-U \cite{peterka_quantification_2024}, NSTX/NSTX-U \cite{ferraro_error_2019,park_sensitivity_2012,menard_progress_2010}, K-STAR \cite{yang_parametric_2021,yang_tailoring_2024,park_overview_2022}, among others. As it is device-agnostic, it has also been used to fit experimental scaling laws \cite{logan_empirical_2020,logan_robustness_2020} for locking thresholds. The combination of a complete EF source model and the empirical scaling laws provides a self-consistent framework for measuring and managing risk in new devices. 

\begin{figure}[h]
\centering{}\includegraphics[width=\linewidth
]{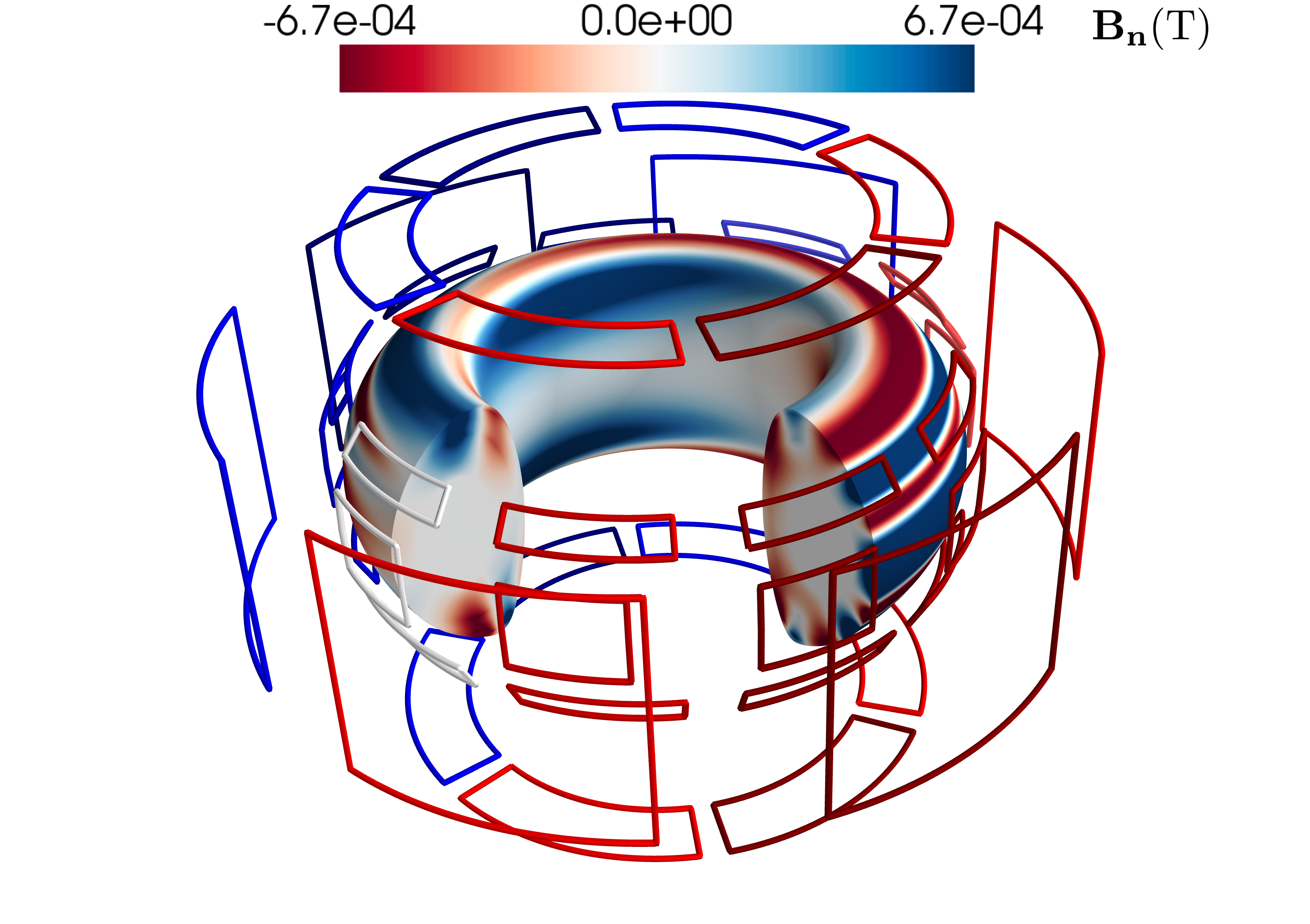}\protect\caption{ITER's available EFC and RMP ELM Coils for error field correction. Coils are shown with $n=1$ toroidally varying currents as will be used for error field correction. Outer coils are EFC coils, inner coils are ELM coils. The field on the plasma shown is from a $n=1$ toroidally varying field applied through an amplitude of 10kAt in each coil set. Control surface fields shown in Tesla. \label{fig:EFCC_plot}}
\end{figure}

Following the success of such coils on many tokamaks, ITER will employ an array of error field correction coils (EFCCs) to destructively interfere with the flux-surface-normal component of the intrinsic error field \cite{foussat_overview_2010}. These are seen in Fig.~\ref{fig:EFCC_plot}. EFCCs can be used to generate a proxy error field for locked-mode studies, and to measure the ideal correction current for regular operation. These measurements are typically known as `compass scans' \cite{hender_effect_1992,la_haye_critical_1992} as they involve scanning the correction current in magnitude and phase and measuring some critical quantity, whether that be a property of the created island \cite{strait_magnetic_2015}, NTV braking \cite{menard_progress_2010, shiraki_error_2014}, or simply forcing the error field to penetrate. The former two methods are non-disruptive, though they require accurate magnetic or angular momentum diagnostics which may not be available on all devices, such as a pilot plant. Forcing the plasma to lock can cause disruption (though this has shown to sometimes be avoidable \cite{paz-soldan_non-disruptive_2022, piron_error_2024, hu_non-disruptive_nodate}) is potentially dangerous depending on the stored energy in the plasma, as a disruption can cause a thermal quench which may damage device components. ITER has notoriously low disruption tolerance due to its unprecedented stored energies, and as such, disruptive error field measurement methods are intolerable. This motivates enabling measurements in low-power plasmas that can be reliably extrapolated to the $Q=10$, 15 MA ITER Baseline Scenario (IBS) \cite{iter_organization_iter_2018} and similar plasmas.

This paper is structured as follows. First, in Section \ref{sec:Model} we will lay out a model based on first order perturbations of the ideal MHD equations and Fourier techniques that robustly models the process of external 3D fields driving a plasma response and thus driving tearing within the plasma; we will also reference an engineering scaling law, which, in combination with the model, can be used to determine the risk of locking. Then, in Section \ref{sec:SourceModel} we will document an error field source model that exhaustively includes all primary device coils and 3D field coils, including the EFCs and ITER ELM Coils (IECs); this concludes with an assessment of ITER's present risk of experiencing locked modes in the IBS. Following this, we describe a new modeling workflow in Section \ref{sec:CompassScans} through which one can test how well an experimental `compass scan' measurement can extrapolate to different plasmas, and argue in favor of measuring the IBS error field source in a low power plasma with coil currents scaled proportionally in unison to those of the IBS. Finally, we conclude in Section \ref{sec:MultiEffects} with a comparison to using physics and precise metrology, and outline advantages and drawbacks of both approaches to EFC.

\section{\label{sec:Model} Model}

\subsection{3D Field Perturbations}

The DCON code \cite{glasser_direct_2016} minimizes the Ideal MHD energy principle with Lagrangian displacements ${\bf \xi}_m$ for a particular toroidal mode number across a spectrum, $m$, of $M$ total poloidal harmonics. These Lagrangian displacements are associated with perturbed magnetic fields ${\bf b}_m = \nabla\times\left({\bf \xi}_m\times {\bf B}\right)$. To carry out this calculation, DCON reduces the MHD force balance problem to a surface layer matrix equation at the edge of the plasma. This surface is known as a control surface. The Generalized Perturbed Equilibrium Code (GPEC) \cite{park_computation_2007, park_shielding_2009, park_importance_2009, logan_neoclassical_2013,park_self-consistent_2017} uses the results of DCON and in turn calculates the linear relationship between a flux derived from vacuum sources and corresponding surface currents. We use Biot-Savart to calculate the half-area weighted vacuum flux as in \cite{park_error_2011,logan_dependence_2016, scoville_multi-mode_2003}, which has the Fourier decomposition,
\begin{equation}
    \tilde{\Phi}^x_{mn} = \frac{1}{(2\pi)^2}\iint {\bf b}^x \cdot \hat n e^{\left(-im\vartheta-in\varphi\right)}\sqrt{\mathcal{J}\left|\nabla \psi\right|} d\varphi d\vartheta.
\end{equation}
\noindent GPEC then finds \cite{park_computation_2007} the linear relationship,

\begin{equation}
    \tilde{\bf \Phi}^r = C \tilde{\bf \Phi}^x,
\end{equation}

\noindent where the components $\tilde \Phi^r_i$ denote the effective (i.e. quantity shielded) resonant flux at the $i$-th rational $q$ surface within the computational domain corresponding to a given toroidal mode number \cite{logan_identification_2016,park_shielding_2009}. $\tilde{\bf \Phi}^x$ is a matrix vector whose components are the $m$-spectrum of the external flux, $\{ \tilde{\Phi}^x_m, m_\text{min} \leq m \leq m_\text{max}\}$, for a given $n$. We refer to $C$ as the resonant coupling matrix. All calculations considered henceforth will be inspecting the $n=1$ physics of ITER. 

This calculation is particularly advantageous compared to previous error field modeling methods, as it includes the effects of poloidal mode coupling, an ideal MHD process by which shielding currents inside the plasma can transform an external perturbation peaked about one poloidal harmonic and drive an entirely different spectrum within the plasma. This is crucial for accurately modeling the error field sensitivity of a given scenario \cite{park_3d_2018, igochine_plasma_2023, paz-soldan_importance_2014, paz-soldan_spectral_2014, wang_toroidal_2020,wang_observation_2016,park_mdc-19_2017}. 

\begin{figure}[h]

    \begin{subfigure}[b]{\linewidth}
        \centering\includegraphics[width=0.97\linewidth]{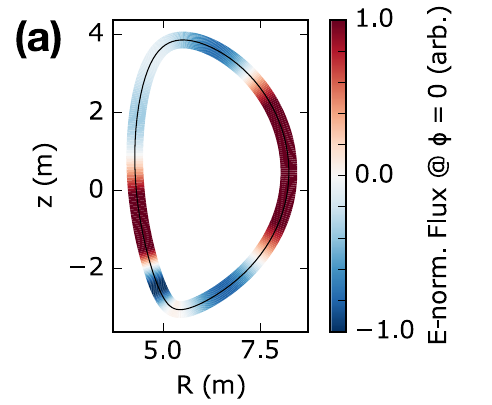}
    \end{subfigure}
    \begin{subfigure}[b]{\linewidth}
        \centering\includegraphics[width=\linewidth]{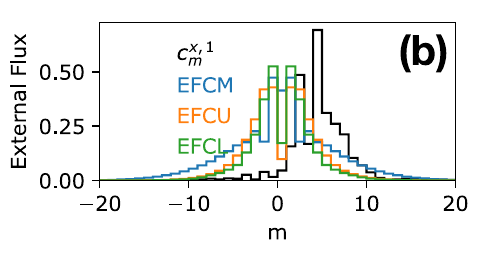}
    \end{subfigure}
    \begin{subfigure}[b]{\linewidth}
        \centering\includegraphics[width=0.99\linewidth]{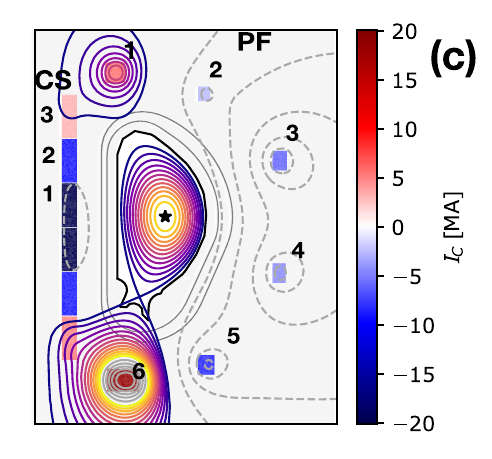}
    \end{subfigure}
\caption{(a) Shows dominant mode on control surface. (b) is the mode's spectral breakdown in PEST magnetic coordinates as well as the spectrum of fields applied by correction coils. (c) shows the IBS with its respective coil currents as calculated by OpenFusionToolkit's TokaMaker \cite{hansen_tokamaker_2024,hansen_hansecopenfusiontoolkit_2024}. Upper CS coils and PF coils labeled. \label{fig:IBS_dommode}}
\end{figure}

We must also define to what we refer as a "dominant mode." It is usually expected that a 2/1 mode will be the most likely to tear, but to generalize we can take a singular value decomposition of the core resonant coupling matrix $C$. Taking an SVD of this core, the right singular vectors (RSVs) (where the number of vectors is equal to the number of rational surfaces) have corresponding singular values which represent the core-resonant effective flux per unit field applied in that spectrum. For the purposes of analyzing locking risk, the domain for surfaces included in the SVD is assumed to be the `core' of the plasma where tearing modes have a high risk of disruption. Here, the `core' refers to the surfaces not in the edge of the plasma, $0 \leq \Psi_N \leq 0.9$, where $\Psi_N$ is the normalized poloidal flux coordinate -- edge surfaces are neglected from this calculation, as the aforementioned scaling laws were developed with a regression to mode agitation on these surfaces only for practical reasons explained in \cite{logan_robustness_2020}. The RSV with the largest singular value is the dominant mode, and likewise the most conducive to mode locking. In order for us to consider solely the dominant mode to quantify locking risk, the singular value of the dominant mode must be much larger than the singular value of all other modes so as to `drown-out' the other modes. In the IBS, there is only one flux surface within the core region, and therefore only one mode, so this does not concern us; however, further study of this singular value separation in ITER can be found in \cite{park_error_2008}. The dominant mode in the IBS is shown in Fig.~\ref{fig:IBS_dommode}. Typically, the dominant mode is most sensitive to low-$m$ perturbations on the low-field side of the plasma; however, given the inclusion of only one rational surface within the core region, the dominant mode is unusually sensitive to high field side perturbations. Fig.~\ref{fig:IBS_dommode} (b) shows the spectral breakdown of the dominant mode, along with those of the applied fields from the EFC coils. It is clear that the correction coils are not perfectly coupled with the dominant mode; any 3D coil set with a spectrum that is non-orthogonal to the dominant mode could fully correct it; however, an ideal coil set would do so with coils that have a high percentage coupling, as this would minimize the non-resonant spectrum being driven by the correction. Both the physics of the dominant mode spectrum and correction coil design are explored extensively in \cite{logan_physics_2021}. To measure this coupling, we will use a quantity called the dominant mode overlap to quantify the precision and amplification with which an external 3D error field spectrum drives the dominant mode. This is defined, normalized by the toroidal field, as, 

\begin{equation}
    \delta \equiv \frac{1}{B_T}{\bf c}^{x,1} \cdot {\bf \Phi}^x,
\end{equation}\label{eq:overlap}

\noindent where $B_T$ is the equilibrium toroidal field and ${\bf c}^{x,1}$ is the singular vector of $C^x$ corresponding to the highest singular value, i.e. the dominant mode. The name here refers to the external flux matrix vector `overlapping' with the first RSV of the resonant coupling matrix -- i.e. the amount the induced non-axisymmetric radial fluxes on the control surface line up with the $m$-spectrum that contributes the most to driving core resonances within the plasma. This model is effective when there is a high separation of singular values, meaning other modes are poorly amplified by the plasma response in comparison. This has shown to be an effective approximation \cite{park_error_2008}. This framework, and the SVD/dominant mode overlap metric in particular, has been employed in error field correction in tokamak experiments and studies for years \cite{park_error_2008,park_error_2011,paz-soldan_spectral_2014,paz-soldan_importance_2014} and is the current metric used by the ITPA for evaluation of error field sensitivities in ITER \cite{park_mdc-19_2017}. These are all naturally complex quantities as external fields can constructively or destructively interfere, so all non-axisymmetric perturbations have a phase. We note that in GPEC simulations, the magnitude of $\tilde \Phi^x_m$ scales linearly with the magnitude of the current in the 3D field sources, and therefore $\delta$ can be considered a measure of the resonant magnetic field for a given coil current; dividing by the coil current will give the resonant overlap $(\delta)$ per kiloampere. In cases where the 3D field source is a small tilt or shift of a nominally axisymmetric coil at fixed current, the quantity can also be considered a measure of the resonant field per millimeter or milliradian. In this paper, we will use $\epsilon$ to denote quantities representing resonant overlap per kiloampere, and $\zeta$ to denote quantities representing resonant overlap per kiloampere per millimeter. Thus, for a perturbation $X$ of coil $j$ with current amplitude $A$,  

\begin{equation}
\delta_j = A \epsilon_j = A X \zeta_j.
\end{equation}

\begin{figure}[t]
    \centering
    \includegraphics[width=1\linewidth]{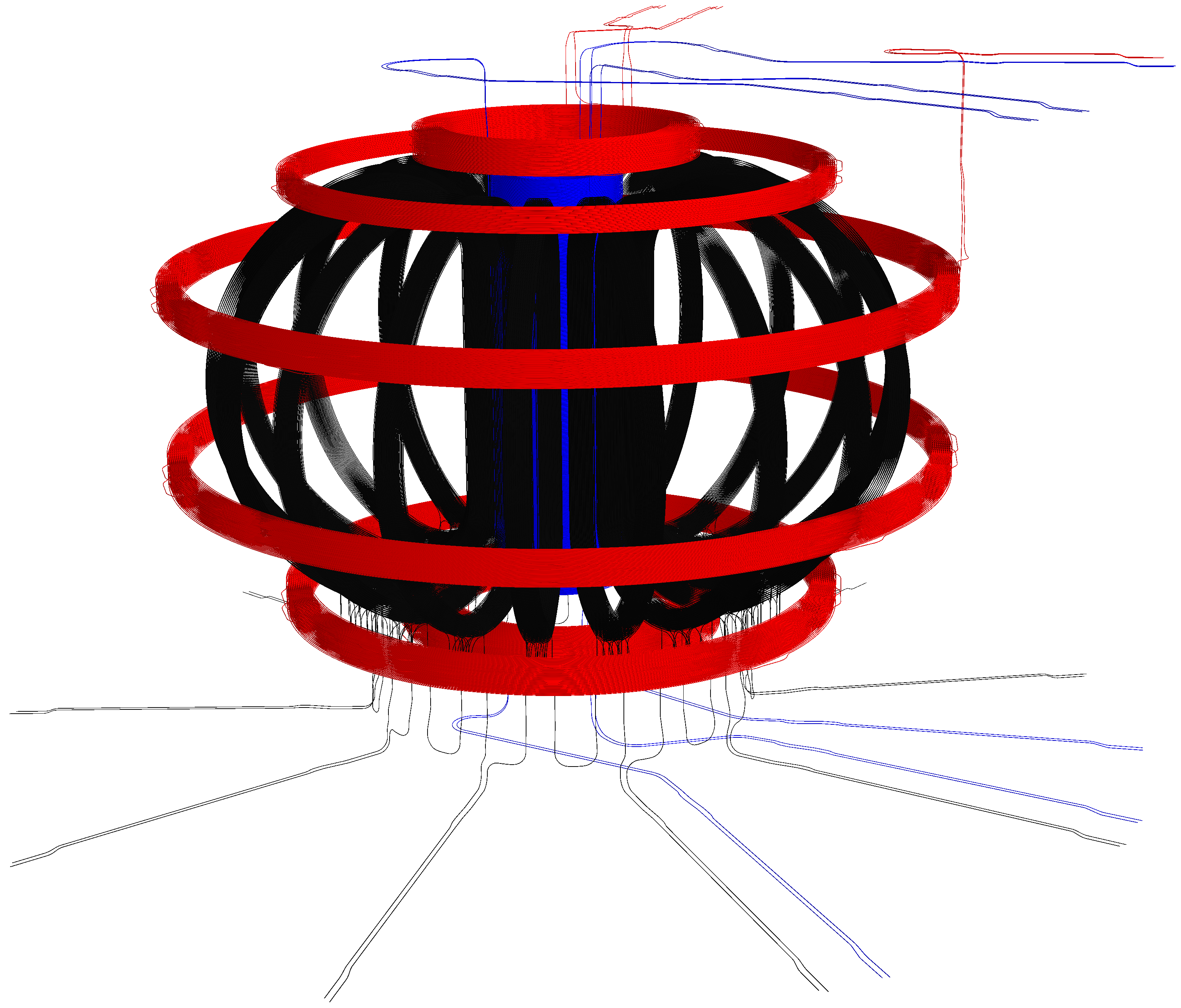}
    \caption{Full 3D Nominal Windings. PF coils in red, TFs in black, CS in blue.}
    \label{fig:nom_windings}
\end{figure}

\subsection{Resonant Error Field Locking Threshold\label{Locking Threshold Model}}

This model of the resonant error field is complimented by empirical locking threshold scaling laws found in \cite{logan_robustness_2020,logan_empirical_2020}. These scaling laws provide an experimentally derived, device-agnostic threshold in $\delta$ for the expected locking flux for a given density, on-axis field, normalized beta, and major radius. They therefore take the form,

\begin{equation}
    \delta_{pen} = 10^{\alpha_{coeff}} n_e^{\alpha_n} B_T^{\alpha_B} R^{\alpha_R} \left(\frac{\beta_N}{\ell_i}\right)^{\alpha_\beta}.
\end{equation}

\noindent As the scaling law we use was regressed using a weighted least squares method (labeled O,L; WLS in \cite{logan_empirical_2020}), $\alpha_i$ have an associated uncertainty; this gives rise to a probability density function for the threshold $\delta_{pen}$ that is near-gaussian and has a width dictated by these uncertainties. The non-gaussian properties of this function arise due to the fact that the uncertainties in the different $\alpha$ can be large, therefore skewing the distribution function. As this regression was performed on a multi-device database with identified locking fields $\delta$ calculated by GPEC, this provides a direct comparison for our source model; the uncertainty contained therein therefore also incorporates any approximations made by the model employed within GPEC.

\section{\label{sec:SourceModel} Error Field Source Model}

\begin{figure}[t]
    \centering \includegraphics[width=0.8\linewidth]{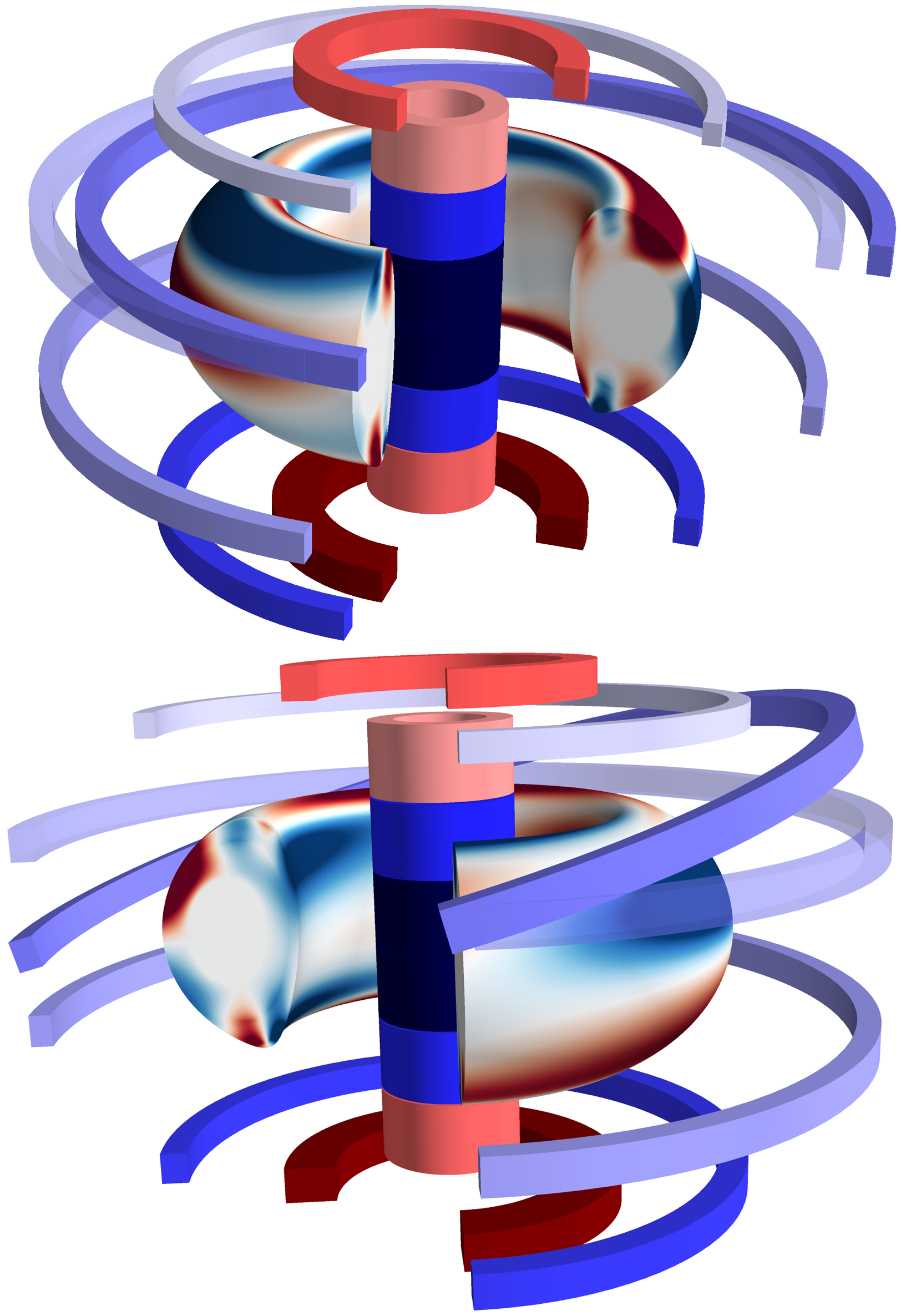}
    \caption{An exaggerated shift (top) and tilt (bottom) in the third Poloidal Field Coil and associated perturbed equilibria. Fields shown are the $b_\text{norm}$ associated with the coil perturbation. Placement tilts/shifts of less than a centimeter are capable of creating large resonant error fields. Transparent coil is nominal placement. Coil is shown as a uniform block for display.}
    \label{fig:pfcs_coils}
\end{figure}

We must develop a model for the error field sources in ITER in order to explore their correction. The primary sources of error fields in tokamaks include the inherent non-axisymmetry of real wound coils (called the `nominal' error field, windings seen in Fig.~\ref{fig:nom_windings}), the nonaxisymmetric coil leads that supply current (Fig.~\ref{fig:nom_windings}), and the misplacement of central solenoid coils, poloidal and toroidal field coils (see Fig.~\ref{fig:pfcs_coils}). In this study, we will consider rigid translational shifts and tilts along the PF and CS coils' principal axes. Work has been done on this in the past \cite{amoskov_optimization_2015, park_mdc-19_2017, amoskov_advanced_2018}, but we look to improve its completeness with complete PF/CS centerline-traced nominal windings and leads, use the newer $\delta$-based model, and conduct a new statistical analysis based on uncertainty intervals in the scaling law to evaluate locking risk. We will also include the final result of previous, robust work on the error field sensitivity of the coherent tilting, shifting, and deformation of the TF coils \cite{mcintosh_iters_2023}. In toroidal field coils, as they are not individually toroidally symmetric, the deformation of the coils can still have significant $n=1$ toroidal harmonic exposure, in contrast with PF/CS coils, whose deformations are primarily $n=2$ and higher. Unlike PF and CS coils, additionally, a simple shift or tilt in a toroidal field coil is not individually prone to having a large $n=1$ exposure; TF coils behave similarly to a delta function about the toroidal angle, and therefore a statistical treatment of coherent toroidal field coil shifting/tilting was built, referenced above. We have included the final toroidal field normalized overlap distribution function in our source model. 

\begin{figure}[h]
    \begin{subfigure}[b]{\linewidth}
        \centering\includegraphics[width=\linewidth
        ]{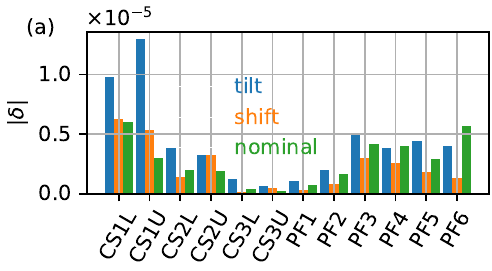}
    \end{subfigure}
        \begin{subfigure}[b]{\linewidth}
            \centering\includegraphics{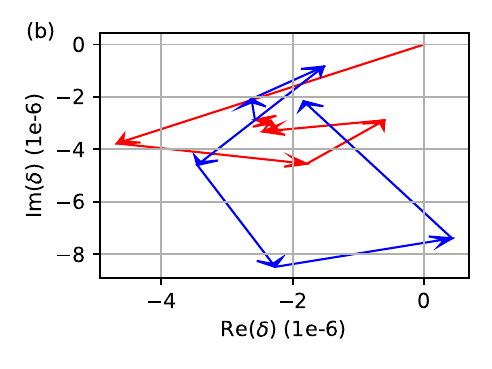}\protect
    \end{subfigure}
    \caption{Coil overlap magnitudes (per mm shift or edge mm tilt) at the IBS's currents as calculated by GPEC. These sources are complex and therefore do not sum directly. Nominal overlaps shown to nearly cancel in (b); CS nominal sources shown in red; PF nominal sources shown in blue.}\label{fig:tiltshift_overlap}
\end{figure}

We see the effect of these 3D field sources in Fig.~\ref{fig:tiltshift_overlap}. The quantity $\delta$ is clearly the highest for the tilting and shifting of PF coils, PF3 and PF4, which are near the center of the low field side of the plasma. The nominal contributions look large to begin with, but as seen in Fig.~\ref{fig:tiltshift_overlap} (b), these sources nearly cancel.

\begin{figure}[h]
\centering{}\includegraphics[width=\linewidth
]{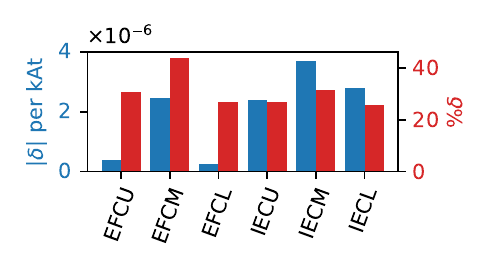}\protect\caption{EFC overlaps with dominant mode. The high degree of overlap between the dominant mode and the EFCM coil is evident. Note, the ITER ELM coils (denoted IEC) also have significant overlap due to their adjacency to the plasma low field side. \label{fig:efc_overlap}}
\end{figure}

Fig.~\ref{fig:efc_overlap} shows the efficacy of using the 3D field coils to correct the dominant mode. The middle EFC coil, EFCM, exhibits the highest percent coupling with the dominant mode; that is to say, its poloidal spectrum lines up the most of all the coils, seen in Fig.~\ref{fig:IBS_dommode} (b).  This is due to its alignment with large features of the dominant mode, in Fig.~\ref{fig:IBS_dommode}. However, due to their proximity to the plasma, the RMP ELM coils get more coupling per kA-turn. We prefer to use the EFCM coil for correction, as the ELM coils have limitations in their total current and have another critical role, suppressing dangerous ELMs. Additionally, the EFCM coil avoids non-resonant mode `pollution', which has implications for the plasma's neoclassical toroidal viscosity. 

\begin{figure}[h]

\begin{subfigure}[b]{\linewidth}
    \centering\includegraphics[width=\linewidth]{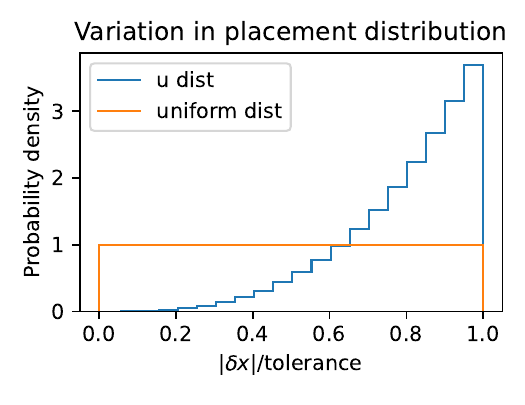}\caption{A pessimistic distribution function peaked at the edge, $P(x)\sim|x|^3$, and an optimistic distribution exhibiting random placement within tolerance.}\label{fig:updf}
\end{subfigure}
\begin{subfigure}[b]{\linewidth}
    \centering\includegraphics[width=\linewidth]{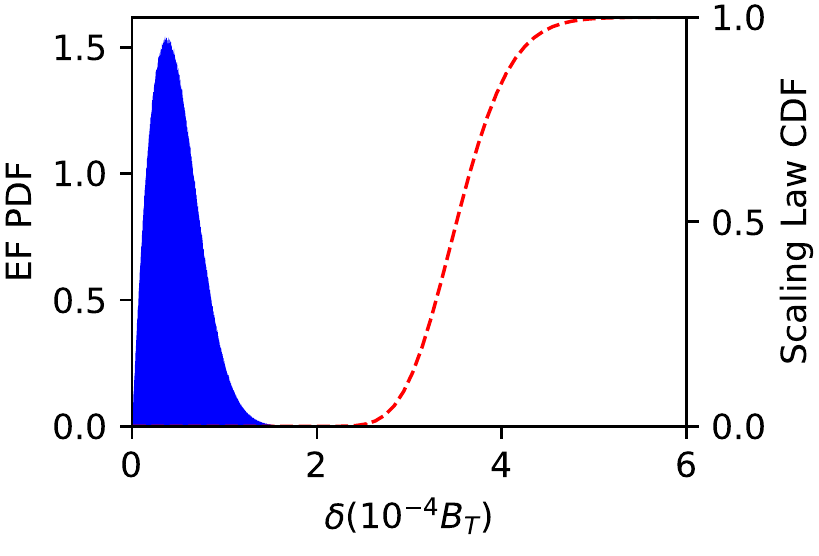}\caption{The PDF for resonant error field in ITER (blue) and the locking threshold probability (red).} 
\end{subfigure}
\caption{Monte Carlo PDF/probabilities for a Monte Carlo simulation using the pessimistic placement distribution function and a total probability curve for the locking threshhold -- probability indicates the chance the threshhold falls below the corresponding $\delta$. \label{fig:mc_pdf}}
\end{figure}
\begin{table}[]
    \centering
    \begin{tabular}{c|c}
         &  $10^4\delta$\\
         Largest possible& 
    2.8\\
 Most likely&0.38\\
 Expectation Value&0.48\\
 Expected Locking Threshold&3.5\\\end{tabular}
    \caption{Various notable values of $\delta$. Notably, all are below ITER's projected locking threshold. The largest possible overlap is the overlap assuming the total nominal overlap and all fields from shifts and tilts, as well as the TF error field, all line up in phase.}
    \label{tab:notable_delta}
\end{table}

To evaluate the error fields to which the IBS is subject, we conduct a numerical Monte Carlo experiment where we assume random tilts and shifts to each PF/CS coil within tolerance. We can then add all the error field sources, some constructively and some destructively interfering, and calculate a total overlap with the dominant mode. The total resonant overlap is given by summing over the $J$ coils,

\begin{equation}
    \delta_{\text{EF}} = \sum_j^J A_j \left( \zeta_{j, \text{tilt}}  \bar \Theta_j + \zeta_{j,\text{shift}}\bar X_j + \epsilon_{j,\text{nom}}\right),
\end{equation}

\noindent where $\zeta_{j,\text{type}}$ is the resonant overlap per kiloampere per millimeter and $\epsilon_{j,\text{nom}}$ is the resonant overlap per kiloampere of the $j$th error field source for each type of source, and $A_j$ is the current in the $j$th error field source. This allows us to scale our error field sources for scenarios with different coil currents. $\bar \Theta$ and $\bar X$ are random variables representing the magnitude of the shift or tilt in millimeters; these are sampled from a distribution function scaled to the tolerance of the given perturbation to the given coil (`increased' tolerances obtained from \cite{amoskov_optimization_2015}). 

As we are working in Fourier space within a $\theta_m, \varphi$ PEST magnetic coordinate system, each quantity is complex to reflect the toroidal phase. We consider in analyzing the Monte Carlo results the \emph{magnitude} of $\delta_{EF}$, but each $\delta$ has a complex component. The random shifts are also complex, the real component representing a shift in the $x$-direction and the imaginary component in the $y$-direction. For additional precision, in the presence of nominal coil windings, one may include both an $x$- and a $y$- shifted and tilted coil, and add these as two separate error field sources. In this case, separate, real $\bar \Theta_j$ and $\bar X_j$ for each direction must be generated, as the phase information will be contained in the $\pi/2$ phase difference between the $x$ and $y$ shifted coils. Options for placement distribution functions for $\bar \Theta_j$ and $\bar X_j$ are shown in Fig.~\ref{fig:updf}. The flat placement distribution model envisions engineers shifting a coil back and forth until it is within tolerance. A more pessimistic model, the polynomial u-distribution, instead pictures engineers slowly lowering a coil into place, and securing it in place the instant it is within tolerance. We chose the more conservative u-curve for the following analysis.

Results of the Monte Carlo experiment are seen in Fig.~\ref{fig:mc_pdf}. The set of sampled $\bar \Theta$ and $\bar X$ for each coil constitutes one Monte-Carlo `universe'. This calculation is repeated for a large number of universes so we may conduct statistics on the probability of ITER being built with a given resonant overlap. Table \ref{tab:notable_delta} shows various important values of $\delta$ to consider: the average over all Monte Carlo universes, the most likely (i.e. peak of the PDF curve), the maximum possible assuming uniform constructive interference in toroidal phase, and the expected IBS locking threshold. We note that all $\delta$ possible fall below the expected locking threshold. 

\begin{figure}[h]
    \centering
    \includegraphics[width=\linewidth]{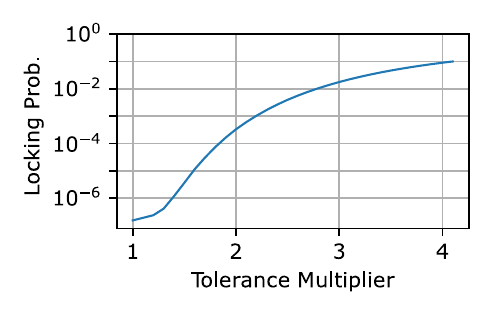}
    \caption{Risk is shown to increase if all tolerances are multiplied by the same coefficient. Tolerances can be increased beyond twice their current values while keeping safety limits of $10^{-3}$ (0.1\% chance of assembly resulting in IBS LMs), showcasing the ITER organization's staunch conservatism on this issue.}
    \label{fig:optimal_tolerance}
\end{figure}

We can also use the aforementioned locking threshold scaling laws to calculate a probability density of locking. As both the threshold and the modeled overlap take the forms of probability density functions, we can doubly integrate to assign a locking risk $r$,

\begin{equation}
    r = \int_0^{\infty} P(\delta) \int_0^\delta R(\delta') d\delta' d\delta,
\end{equation}

\noindent where $R(\delta)$ is the PDF of the locking threshold, whose integral is shown in red in Fig.~\ref{fig:mc_pdf}, and $P(\delta)$ is the probability density function of the resonant mode overlap, shown in blue in Fig.~\ref{fig:mc_pdf}. This figure shows Monte Carlo simulations of an uncorrected IBS based on the EF source model. The risk, when integrated, is found to be $1.55\times10^{-7}$, indicating that ITER's coil tolerances as a whole are extremely conservative. Fig.~\ref{fig:optimal_tolerance} shows the result of raising all tolerances by a uniform multiple, indicating that tolerances could be increased by a factor of approximately $2.2$ and still preserve near-zero $10^{-3}$ probability of locking in the IBS, without beginning to consider error field correction. Allowing an assumption of functional error field correction, tolerances could be raised even further. Further work in this paper will be using tolerances $2.2$ times higher than prescribed, as these leaves the locking risk at a higher $10^{-3}$. In the case with prescribed tolerances, since every $\delta$ possible falls below the expected locking threshold, the only probability of locking comes from the least squares uncertainty in the scaling law exponents.

\begin{figure}[t]
    \includegraphics[width=\linewidth]{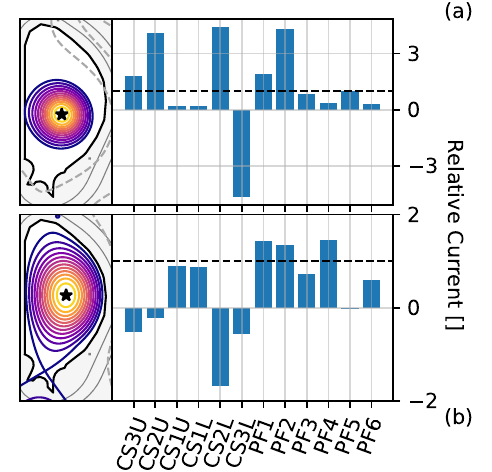}
        \caption{Low-power early ITER equilibria can be seen here (a) LL and (b) LH, alongside their scaled coil currents, $\frac{I_{coil}}{I_{coil}^{IBS}}\frac{15 MA}{I_P}$. Relative magnitude and direction of coil currents is crucial to consider in measuring error field via compass scans as equilibria that require radically different proportional currents will offer an entirely different error field source, as different coils have different placement errors. \label{fig:given_eq}}
\end{figure}
\begin{figure}[t]
    \includegraphics[width=\linewidth]{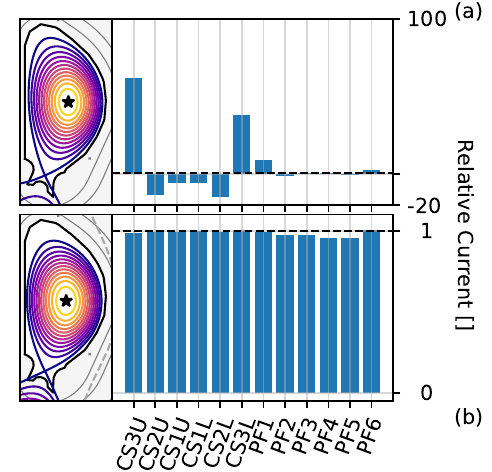}
     \protect\caption{(a) Scenaro MS; (b) Scenario MC. New Tokamaker Scenarios either constraining shape or relative coil currents. Relative coil currents $\frac{I_{coil}}{I_{coil}^{IBS}}\frac{15 MA}{I_P}$ will dictate the similarity of the 3D field source across plasmas. \label{fig:created_scen}}
\end{figure}

\section{\label{sec:CompassScans} Conducting accurate compass scan measurements}

Compass scans are a standard measurement technique for observing apparent error fields and optimizing their correction. The simplest form of compass scan involves ramping the $n=1$ current in an array of correction coils continuously until mode locking occurs. This is repeated with different phases of the current in the correction coils, until a circle is drawn in the complex plane of coil phase/magnitude. This circle is fit with a least squares or other method, and the center is taken as the optimal correction current. However, as is central to the premise of this work, mode locking is prone to causing disruptions and therefore must be avoided in plasma shots with a high quantity of stored energy. This applies for ITER, and any proposed pilot plant, which will need large quantities of stored energy in the pursuit of fusion power gain. Therefore, we must consider how to measure the resonant error field, without mode locking in high-energy scenarios. This section presents a method for optimizing compass scan extrapolation from low power scenarios to complement the plethora of methods presented in Sec.~\ref{sec:Introduction}.

We propose designing scenarios that operate with low $I_p$ and $B_T$, and possibly which are shaped fairly differently from a scenario that we wish to correct, but which hold the currents in PF and CS coils proportional in magnitude relative to each other in comparison with the high-power scenario. The scenario we will use as a reference for correction is the IBS. We will therefore compare currently proposed early ITER scenarios (see Fig.~\ref{fig:given_eq}) to both a scenario where shape matching was prioritized but with different coil currents proportional to each other and a scenario with similar coil currents but a different shape, in their effectiveness in determining the optimal error field correction of the IBS. The reason for picking a scenario that proportionally matches coil currents is fairly simple -- we wish to preserve the error field source, a majority of which is expected to come from the PF and CS coils, both in their nominal non-axisymmetry, and their slight placement error due to engineering realities.  

Once we have a Monte Carlo method for evaluating the error field in a given equilibrium, we may simulate the process of completing a compass scan. To do this, we consider what this would look like in each Monte Carlo `universe,' defined by a unique set of tilts/shifts and a unique value of the TF coil error field. We do the following calculations in \emph{each} universe, which will create a new distribution of the different total $\delta$ in each universe. This will enable a statistical assessment of a compass scan measurement. As compass scans are conducted in the complex EFC current phase space, we proceed by defining an exact, complex-valued current amplitude $A^{M,\text{exact}}_\text{EFC}$, which would be measured as the optimal error field correction current in the EFCM coil in an experiment with zero uncertainty. This is defined as, 

\begin{equation}
    A_\text{EFC}^{M,\text{exact}} = \frac{\delta^M_\text{EF}}{\epsilon^M_\text{EFCM}},
\end{equation}

\noindent where the $M$ superscript designates its association with the measurement scenario, $\delta^M_\text{EF}$ is the total resonant overlap for that scenario, and $\epsilon^M_\text{EFCM}$ is the resonant overlap per kiloampere associated with the EFCM correction coil. In order to assign a measurement uncertainty, we sample points from a random distribution centered at $A^{M,\text{exact}}_\text{EFC}$. The type of distribution used here can significantly affect what qualifies as a tolerable `uncertainty'; we will discuss this later. This is carried out as,

\begin{equation}
    A^{M}_\text{EFC} = A^{M,\text{exact}}_\text{EFC} \left(1 + \bar Y\exp\left(i\bar \Theta\right)\right),
\end{equation}\label{eq:currentuncert}

\noindent where $\bar \Theta$ is sampled from a uniform phase distribution $\left[0,2\pi\right)$ and $\bar Y$ from the aforementioned uncertainty distribution.

From here, the residual resonant error field after applying the measured current $A^{M}_\text{EFC}$ in the EFCM correction coil can be calculated, including a scaling by plasma current across different scenarios:

\begin{align}
    \delta^\text{IBS}_\text{residual} &= \delta_\text{EF}^\text{IBS} - A_\text{EFC}^M\epsilon_\text{EFCM}^\text{IBS}\frac{I^\text{IBS}_P}{I_P^M}\\
    &= \delta_\text{EF}^\text{IBS} - \delta_\text{EF}^M \frac{\epsilon_\text{EFCM}^\text{IBS}}{\epsilon_\text{EFCM}^M}\frac{I_P^\text{IBS}}{I_P^M} \left(1 + \bar Y\exp\left(i\bar \Theta\right)\right).
\end{align}

\noindent This form exposes the two approximations that are often assumed in employing compass scans. As $\delta_\text{residual}^\text{IBS}$ is to be minimized, we are assuming that the EFCM coupling to each scenario is similar so that  $\epsilon_\text{EFCM}^\text{IBS}/\epsilon_\text{EFCM}^M \approx 1$, and that $\delta_\text{EF}$ maintains its phase across scenarios and scales with $I_P$. The first criterion has found to be somewhat accurate in ITER plasmas \cite{park_mdc-19_2017, park_error_2008}, though this should be tested when considering extrapolability on other devices with a code like GPEC. The second criterion is true in some cases but not all. DIII-D employs a slightly more complex scaling, but the conclusion is the same; for optimal error field correction, a measurement scenario should be designed such that $\delta_\text{EF}^M I_P^\text{IBS}/\delta_\text{EF}^\text{IBS}I_P^M\approx 1$ (or with whichever scaling is found to be effective). 

Enabled with this method, we will test the measurement efficacy two of planned scenarios, LL and LH, and two new scenarios, MS, and MC, against the IBS. LL is an early 2 MA L-mode ITER plasma with no shaping, and LH is a later 5 MA H-mode plasma with IBS-like shaping \cite{iter_organization_iter_2018}, seen in Fig.~\ref{fig:given_eq}. MS is a 2.4 MA L-mode scenario constrained in shape but choosing coil currents disproportionate to those of ITER, and MC is an L-mode scenario designed to match coil currents proportionally based on a scaling of $I_P$. These are shown with their respective coil currents in Fig.~\ref{fig:created_scen}. Due to the limitation that the toroidal field can only take on full, half, and third-field values, and given the toroidal field's huge contribution to the total resonant overlap, this means that we choose a 5 MA plasma to match the toroidal field for MC which has a third-field toroidal field.

\begin{figure}[h]
\centering{}\includegraphics[width=0.99\linewidth
]{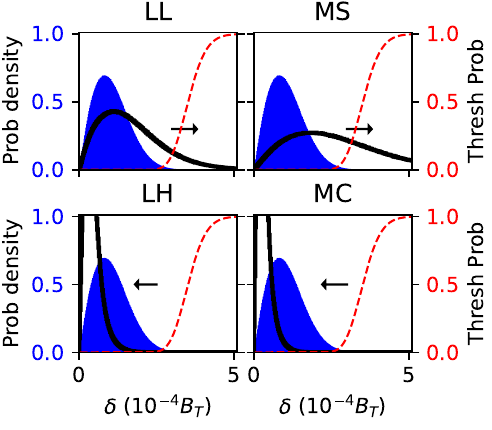}\protect\caption{In a compass scan Monte Carlo, the un-corrected, tolerance-boosted (tolerances multiplied by 2.2 to uncorrected $10^{-3}$ risk factor) $\delta$ distribution (blue) is shifted based on the effects of error field correction to a new $\delta^\text{IBS}_\text{residual}$ distribution (black) either towards or away from the locking threshold (red). \label{fig:mc_compass}}
\end{figure}

Distributions of $\delta^\text{IBS}_\text{residual}$ are seen in Fig.~\ref{fig:mc_compass} for comparison with the distribution of $\delta_\text{EF}$ for the IBS. We chose four scenarios for comparison, as viewed in Table~\ref{tab:scenarios}. The Low power L mode (LL) is akin to an early first stage plasma soon after breakdown is first achieved, within the Start of Research Operation stage. The Low power H mode (LH) is an H-mode scenario from far later in the same campaign. The first new case (MS) a scenario constrained in shape but choosing coil currents and a toroidal field  disproportionate to those of the IBS. The second new case (MC) is a scenario designed to match coil currents proportionally based on a scaling of $I_P$. These new scenarios were created for this error field study using TokaMaker \cite{hansen_tokamaker_2024,hansen_hansecopenfusiontoolkit_2024}.

\begin{figure}[h]
\centering{}\includegraphics[width=\linewidth
]{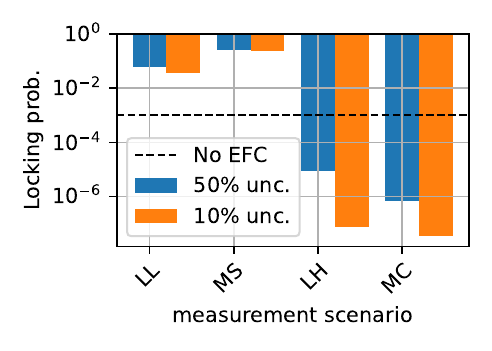}\protect\caption{Risks are shown for the IBS using correction currents measured in each scenario, both at 10\% and 50\% assumed measurement uncertainty. Y-axis is log scale; lower is more effective EFC. \label{fig:risks} }
\end{figure}

Fig.~\ref{fig:risks} displays the distribution of $\delta^\text{IBS}_\text{residual}$ integrated with the locking distribution, giving us a risk for the IBS using an EFC scheme derived from a compass scan in the labeled scenario. We see that using the MC scenario to conduct a compass scan significantly collapses the resonant overlap distribution leftward and massively reduces locking probability from the risk of running the IBS with no error field correction.  The locked mode risk is reduced from $10^{-3}$ to $5.0\times10^{-7}$ through a correction measured in MC. This stands in contrast to scenario MS with a similar shape but different PF 2, 3, and 4 currents, i.e. a completely different error field source; if we use this scenario to measure the resonant error field for the IBS, using EFCM on this basis actually \textit{increases} locking risk. We also find that the LL scenario suffered due to not being a 5 MA (one third of IBS current) scenario while having a one-third-field toroidal field. The worst case scenario would then be a scenario with PF/CS currents proportional to those in the IBS but opposite in direction and $I_P$ far less than 5 MA. Effective error field source extrapolation can be observed as well in the highest power non-IBS scenario we tested; this scenario had a somewhat similar error field source, as all the highest contributors to $\delta$ have similarly signed currents (Fig.~\ref{fig:given_eq}).

\begin{table*}[t] 
    \centering
    \begin{tabular}{|l|lllllllll|}
    \hline
    Alias & $I_P$   & $B_T$  & $\kappa$ & $\delta$ & $\beta_P$ & $U_\text{plasma}$ & $q_{95}$ & Source    & Equilibrium Type \\ \hline
    LL    & 2 MA& 1.76 T & 1.0      & 0        & 3.3 \%    & 380 kJ            & 2.58     & DINA      & Planned Scenario   \\
    LH    & 5 MA    & 2.65 T & 1.84     & 0.45     & 3.66 \%   & 2.28 MJ           & 4.34     & DINA      & Planned Scenario   \\
    MC& 5 MA& 1.76 T & 1.75& 0.40& 23.2 \%& 14.9 MJ& 2.75& TokaMaker & Currents Matched\\
    MS& 2.4 MA& 1.76 T & 1.86& 0.47& 48.3 \%& 7.02 MJ& 6.19& TokaMaker & Shape Matched   \\
    IBS   & 15 MA   & 5.3 T  & 1.86     & 0.47     & 53.1\%    & 298 MJ            & 2.89     & DINA      & Planned Scenario           \\ \hline
    \end{tabular}
    \caption{Summary of scenarios shown in Figs.~\ref{fig:given_eq} and \ref{fig:created_scen}. Note the extreme difference in stored energy in the IBS v.s. the selected scenarios. Scenario MC was designed to proportionally match the coil currents used in the IBS while decreasing the stored energy. \label{tab:scenarios}}
\end{table*}

\section{\label{sec:MultiEffects} Comparisons with a model-based approach}

The use of GPEC and its supporting codes is clearly an effective tool for studying error field physics in tokamaks. Current and future devices such as ITER have or have planned excellent metrology \cite{kim_study_2003,kim_status_2005,shibanuma_assembly_2013,wenge_alignment_2005,braeuer_metrology_2006,pedersen_confirmation_2016} that can help determine coil placement errors such that an error field source model may be built for the constructed device without compass scan measurements. We are therefore interested in evaluating the advantages and disadvantages of each approach as well as the overall effectiveness at preventing error field penetration.

To evaluate this, we will perform an additional Monte Carlo calculation in which we add random complex-valued uncertainty to each $\bar X_i$ and $\bar \Theta_i$ and recalculate the total resonant 3D field, which this time is the resonant error field we would expect to calculate using a code like GPEC; we denote this quantity $\delta_\text{EF,calc}$. Then, the correction current an operator of the device would apply based on the model is simply, 

\begin{equation}
    A_\text{EFC}^\text{IBS calc} = \frac{\delta_\text{EF}^\text{IBS met}}{\epsilon_\text{EFCM}^\text{IBS}}.
\end{equation}

\noindent where $\epsilon_\text{EFCM}^\text{IBS}$ is as before and $\delta_\text{EF}^\text{IBS met}$ is the total resonant overlap as calculated by GPEC; included in this is uncertainty from error in metrology. This current $A_\text{EFC}^\text{IBS calc}$ is not assigned an additional uncertainty as we assume all physics inaccuracies in the model are integrated into the uncertainties in the scaling laws, as described in \cite{logan_robustness_2020}.  We will assume a range of values for this uncertainty to inspect sensitivity. The residual overlap is now, 

\begin{equation}
    \delta_\text{residual}^\text{IBS} = \delta_{EF}^\text{IBS} - \frac{\delta_\text{EF}^\text{IBS met}}{\epsilon_\text{EFCM}^\text{IBS}}\epsilon_\text{EFCM}^\text{IBS} = \delta_{EF}^\text{IBS} - \delta_\text{EF}^\text{IBS met}.
\end{equation}

\noindent The risk of locking, here, is now defined as the distribution of $\delta_\text{residual}^\text{IBS}$ integrated with the locking probability.

\begin{figure}[h]
\centering{}\includegraphics[width=\linewidth
]{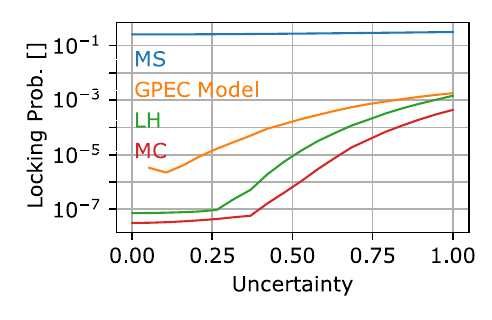}\protect\caption{Curves showing the relationship between risk of locking while using a given error field correction strategy versus the expected accuracy in compass scan or metrology measurements. For metrology and GPEC, the $x$-axis refers to the relative uncertainty in metrology measurements. A value is omitted at zero uncertainty for the model curve as risk is definitively zero.\label{fig:modelbasedexp}}
\end{figure}

Many future devices plan to have a very detailed metrological documentation of their devices for error field source modeling among other uses -- this makes it pertinent to discuss the applicability of compass scans in the presence of excellent metrology. We address this with a statistical experiment in which we scan an assumed random gaussian error in the compass scan centroid. We also do a similar set of calculations varying a percent error in the measured tilts and shifts of coils. Fig.~\ref{fig:modelbasedexp} shows the probability of locking based on the Monte Carlo simulations of compass scans as well as the additional Monte Carlo simulation of metrology error. We note that there is no reason to consider these two separate forms of measurement on the same scale of uncertainty. The metrology process itself is expected to have far lower than 50\% measurement uncertainty, though coils move within their casings via thermal expansion and $J\times B$ forces, which can raise this uncertainty significantly.

There are multiple takeaways from this result. First, we conclude definitively that conducting EFC with currents measured through a compass scan in the wrong scenario can be disastrous, as seen from MS. Second, GPEC is shown to be very effective even with a high degree of metrology error; even getting the general `direction' (measured position randomly sampled up to the entire tolerance away from its true location, corresponding to the right end of Fig.~\ref{fig:modelbasedexp}) can provide a reduction in resonant error field, though less effective than a compass scan with a reasonable \cite{buttery_limits_2012,schaffer_iter_2011} 50\% measurement uncertainty. We also see the robust advantage of the specially designed MC scenario, save for extremely low uncertainty. This leads to the conclusion that an exercise on ITER using both the metrology data and GPEC as well as a compass scan in a use-tailored plasma in which one tests for agreement would be able to set a very reliable baseline for correction currents in IBS-like scenarios with low risk to the device.

Fig.~\ref{fig:modelbasedexp} tells us two important things; that accurate metrology is often superior to a compass scan with \~50\% assumed uncertainty (assuming coil positions are able to be accurately measured), and that the scenario tailoring strategy of matching currents can still provide robust and reliable results, even in the context of moderate uncertainty.

\section{\label{sec:Conclusion}Conclusion}

3D Field physics will be a recurring topic in future research as progress is made towards achieving burning plasmas in toroidal fusion devices and the prospect of a pilot plant grows nearer. Utilizing our probability distribution function approach to both the resonant overlap and the scaling law, we find that existing ITER tolerances have $1.5\times10^{-7}$ chance of locking in the IBS with no error field correction. Existing tolerances are shown to be approximately $2.2$ times tighter than necessary to achieve $99.9\%$ certainty of no locked modes in the IBS. We also find that using compass scan data from newly proposed equilibrium MC, we can both increase tolerances uniformly by a factor of $2.2$ and simultaneously keep locking probability to $5.0\times10^{-7}$. Further, with error field correction from the MC equilibrium, tolerances could be increased by a factor of about 3.4 while keeping locking probability to $10^{-3}$. The use of metrology even with a wide range of measurement uncertainties is validated as an effective approach to error field correction. Our complete error field source model of ITER and associated locking risk analysis has shown that while engineering tolerances in ITER are more than sufficient for adequately controlling risk of disruption due to error fields, such models must be paid attention to closely when designing new toroidal fusion systems. Experiments and future energy-producing tokamaks built by industry have great interest in cost-saving through relaxing engineering constraints; such a risk analysis workflow is crucial in managing the upper bounds of coil design and manufacturing error in a context where resonant mode overlap is allowed to be higher. The coils for ITER were designed before this analysis was developed, and therefore were constructed and evaluated using an older model; however, this analysis is expected to hasten/cheapen design and construction of coils in the future on devices such as SPARC \cite{sweeney_mhd_2020, creely_overview_2020}.

In ITER and other future experiments, the analysis we have presented in section \ref{sec:CompassScans} is likely to be an effective tool to get more reliable error field correction information from a crude compass scan. Since a small change in the total $m$-spectrum of radial fluxes on the plasma exterior can dramatically alter the 3D field coupling to the dominant mode, taking care in conducting a compass scan in a scenario where the error field source is controlled to be similar to that of scenarios in which one hopes to employ this correction measurement is both important and effective. We have also shown that paying no mind to the error field source for the measurement scenario when measuring optimal correction currents can lead to counter-productive correction fields; if a correction current is measured in scenario MS, applying EFCC counter-intuitively increases the error field. While more accurate methods exist of determining the optimal correction current for a given plasma, these techniques require specific and accurate diagnostics and metrology that are not present on all current devices, and will likely not be present in a fusion pilot plant. Nevertheless, error field correction will remain a concern for managing disruption risk and other unwanted 3D field effects. For that, our source-matching technique is an attractive and practical option for measuring optimal correction currents in low-power plasmas to be used in correcting high stakes plasmas. 

\section{\label{sec:Acknowldgements}Acknowledgements}

This work was supported by the U.S. Department of Energy Office of Science Office of Fusion Energy Sciences under Award DE-SC0022272, and the Technology Development Projects for Leading Nuclear Fusion through the National Research Foundation of Korea (NRF) funded by the Ministry of Science and ICT (No. RS-2024-00350293). We also thank Simon McIntosh of the ITER organization for contributing the Monte Carlo model of the Toroidal Field Coil error fields. 

\printbibliography

\end{document}